\documentclass[12pt]{article}
\usepackage{graphicx,amsmath}
\usepackage{units}

\newlength{\dinwidth}
\newlength{\dinmargin}
\def\lapproxeq{\lower .7ex\hbox{$\;\stackrel{\textstyle                                                    
<}{\sim}\;$}}                                                    
\def\gapproxeq{\lower .7ex\hbox{$\;\stackrel{\textstyle                                                    
>}{\sim}\;$}}                                                    
\def\be{\begin{equation}}                                                    
\def\ee{\end{equation}}                                                    
\def\bea{\begin{eqnarray}}                                                    
\def\eea{\end{eqnarray}}

\def\sh{\hat s}
\def\sh2{{\hat s}^2}

\begin{document}
\titlepage                                                    
\begin{flushright}                                                    
IPPP/11/59  \\
DCPT/11/118 \\                                                    
\today \\                                                    
\end{flushright} 
\vspace*{0.5cm}
\begin{center}                                                    
{\Large \bf From hard to soft high-energy $pp$ interactions\footnote{To be published in the Proceedings of the Low $x$ Workshop, Santiago de Compostela, Spain, 2-7 June 2011.}}\\

\vspace*{1cm}
                                                   
\underline{A.D. Martin}$^a$, M.G. Ryskin$^{a,b}$ and V.A. Khoze$^{a,b}$ \\                                                    
                                                   
\vspace*{0.5cm}                                                    
$^a$ Institute for Particle Physics Phenomenology, University of Durham, Durham, DH1 3LE \\                                                   
$^b$ Petersburg Nuclear Physics Institute, Gatchina, St.~Petersburg, 188300, Russia

\vspace*{1cm}

\begin{abstract}
\noindent We discuss how the main features of high-energy `soft' and `semihard' $pp$ collisions may be described in terms of parton cascades and multi-Pomeron exchange. The interaction between Pomerons produces an effective infrared cutoff, $k_{\rm sat}$, by the absorption of low $k_t$ partons. This provides the possibility of extending the parton approach, used for `hard' processes, to also describe high-energy soft and semihard interactions. We outline a model which incorporates these features. Finally, we discuss what the most recent LHC measurements in the soft domain imply for the model.
\end{abstract}

\vspace*{0.5cm}                                                    
                                                    
\end{center}    

\section{A unified description?}

`Soft' and `hard' high-energy $pp$ interactions are usually described in different ways.  The appropriate formalism for high-energy soft interactions is based on Reggeon Field Theory with a phenomenological (soft) Pomeron, whereas for hard interactions we use a QCD partonic approach, where the (QCD) Pomeron is associated with the BFKL vacuum singularity. However, the two approaches appear to merge naturally into one another.  That is, the partonic approach seems to extend smoothly into the soft domain. 

The BFKL equation describes the development of the gluon shower as the momentum fraction, $x$, of the proton carried by the gluon decreases.  That is, the evolution parameter is ln$(1/x)$, rather than the ln$k_t^2$ evolution of the DGLAP equation. 
Formally, to justify the use of perturbative QCD, the BFKL equation should be written for gluons with sufficiently large $k_t$. However, it turns out that, after accounting for NLO corrections and performing an all-order resummation of the main higher-order contributions, the intercept of the BFKL Pomeron depends only weakly on the scale. The intercept is found to be $\Delta \equiv \alpha_P(0)-1 \sim 0.3$ over a large interval of $k_t$
{\cite{bfklresum}.
%\cite{nll}.
Thus the BFKL Pomeron is a natural object to continue from the `hard' domain into the `soft' region.

The BFKL or QCD Pomeron may be viewed as a sum of ladders based on the exchange of two $t$-channel (Reggeized) gluons. Each ladder produces a gluon cascade which develops in ln$(1/x)$ space, and which is not strongly ordered in $k_t$. There are phenomenological arguments (such as the small slope of the Pomeron trajectory, the success of the additive quark model relations, etc.) which indicate that the size of an individual Pomeron is relatively small as compared to the size of a proton or pion etc. Thus we may regard the cascade as a small-size `hot-spot' inside the colliding protons.

At LHC energies the interval of BFKL ln$(1/x)$ evolution is much larger than that for DGLAP ln$k_t^2$ evolution. Moreover, the data already give hints that we need contributions not ordered in $k_t$, $\grave{a}~ la$ BFKL, since typically DGLAP overestimates the observed $\langle k_t \rangle$ and underestimates the mean multiplicity \cite{cms,atlas}. Further,  it is not enough to have only one Pomeron ladder exchanged; we need to include multi-Pomeron exchanges.

Basically, the picture is as follows. In the perturbative domain we have a single bare `hard' Pomeron exchanged with a trajectory $\alpha_P^{\rm bare}\simeq 1.3+\alpha'_{\rm bare}t$, where $\alpha'_{\rm bare} \lapproxeq 0.05$ GeV$^{-2}$. The transition to the soft region is accompanied by absorptive multi-Pomeron effects, such that an {\it effective} `soft' Pomeron may be approximated by a linear trajectory $\alpha^{\rm eff}_P \simeq 1.08+0.25t$ in the {\it limited} energy range up to Tevatron energies \cite{DL}.  This smooth transition from hard to soft  is well illustrated by the behaviour of the data for vector meson ($V=\rho, \omega, \phi, J/\psi$) production at HERA, $\gamma^*p\to V(M)p$, as $Q^2+M^2$ decreases from about 50 GeV$^2$ to zero.

\section{Multi-Pomeron diagrams}

The {\it eikonal} model accounts for the multiple rescattering of the incoming fast particles. We have\footnote{To allow for low-mass proton dissociation, the amplitude (\ref{eq:eik}) is written in matrix form, $T_{ik}$, between (Good-Walker) diffractive eigenstates.}
\begin{equation}
{\rm Im}T\;=\;(1-e^{-\Omega/2})\;=\;(\Omega/2)-(\Omega^2/8)+...
\label{eq:eik}
\end{equation} 
which display the multi-Pomeron corrections to the bare Pomeron amplitude, $\Omega/2$, that
 tame the power 
growth of the cross section with energy. Experimentally, 
these multi-Pomeron diagrams also explain the growth of the central 
plateau
\cite{cms,atlas}
\be
\frac{dN}{d\eta}~=~n_P \frac{dN_{\rm 1-Pom}}{d\eta},
\label{eq:nP}
\ee
where $dN_{\rm 1-Pom}/d\eta$ is the plateau due to the exchange of one 
Pomeron, which is independent of collider energy. The growth is due to the 
increasing number, $n_P$, of Pomerons exchanged as energy increases. These 
(eikonal) multi-Pomeron contributions are included in the present Monte 
Carlos to some extent, as a Multiple Interaction (MI)
option, but Pomeron-Pomeron interactions are not allowed for. 

Since the (small size) Pomeron cascades (hot spots) occur at 
different impact parameters, $b$, there is practically no interference between them. Moreover, at this `eikonal' stage, 
the multi-Pomeron vertices, which account for the interaction between 
Pomerons, are not yet included in the formalism. These are interactions between partons within an individual hot spot (Pomeron). Formally, these are NNLO interactions, but their contribution is {\it enhanced} by the large multiplicity of partons within a high-energy cascade.  In terms of Reggeon Field Theory, the additional interactions are described by so-called {\it enhanced} multi-Pomeron diagrams, whose contributions are controlled by triple-Pomeron (and more complicated multi-Pomeron) couplings\footnote{These diagrams are responsible for high-mass proton dissociation.}. Recall that non-enhanced (eikonal) multi-Pomeron interactions are caused mainly by Pomerons occurring at different impact parameters, and well separated from each other in the $b$-plane. On the other hand, the enhanced contributions mainly correspond to additional interactions (absorption) within an {\it individual} hot spot, but with the partons well separated in rapidity.

The main effect of the enhanced contribution is the absorption of low $k_t$ partons. Note that the probability of these additional interactions is proportional to $\sigma_{\rm abs} \sim 1/k_t^2$, and their main qualitative effect is to induce a splitting of low $k_t$ partons into a pair of partons each with lower $x$, but larger $k_t$. Effectively this produces a dynamical infrared cut-off, $k_{\rm sat}$, on $k_t$, and partly restores a DGLAP-like $k_t$-ordering within the cascade at larger $k_t$.

\section{Schematic sketches of the model}

Qualitatively, the structure of soft interactions based on the `BFKL' multi-Pomeron approach is as follows. The evolution 
%in ln$(1/x)$ space 
 produces a parton cascade which occupies a relatively small domain in $b$-space, as compared to the size of the proton. We have called this a hot spot. The multiplicity of partons grows as $x^{-\Delta}$, while the $k_t$'s of the partons are not strongly ordered and depend weakly on ln$s$. Recall $\Delta\equiv \alpha_P(0)-1$. Allowing for the running of $\alpha_s$, the partons tend to drift to lower $k_t$ where the coupling is larger. This is shown schematically in Fig.~\ref{fig:abc}(a). 
\begin{figure} 
\begin{center}
\includegraphics[height=4.5cm]{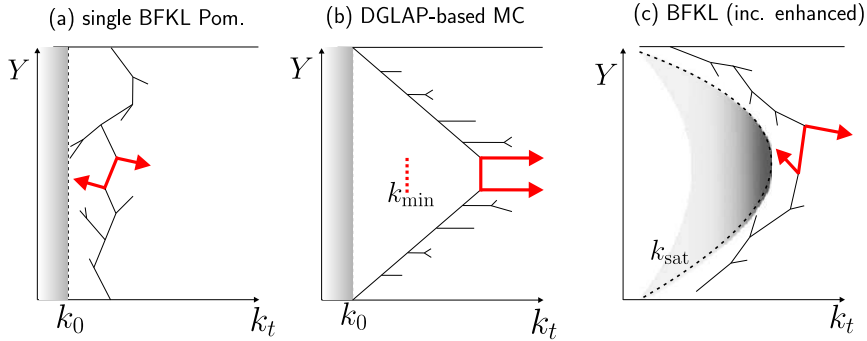}
\caption{\sf Sketches of the basic diagram for semi-hard particle production in $pp$ collisions. The figure is taken from Ref. \cite{KMRmp}.}
\label{fig:abc}
\end{center}
\end{figure}

On the contrary, the DGLAP-based Monte Carlos generate parton cascades strongly ordered in $k_t$. That is, the parton $k_t$ increases as we evolve from the input PDF of the proton to the matrix element of the hard subprocess, which occurs near the centre of the rapidity interval, Fig.~\ref{fig:abc}(b).  Since the cross section of the hard subprocess behaves as $d\hat{\sigma}/dk_t^2 \propto 1/k_t^4$, the dominant contributions come from near the lower limit $k_{\rm min}$, of the $k_t$ integration. In  fact, in order to describe the high-energy collider data, it is necessary to artificially introduce an energy dependent infrared cutoff; $k_{\rm min} \propto s^a$ with $a \sim 0.12$ \cite{P81}. This cutoff is only applied to the hard matrix element, whereas in the evolution of the parton cascade a constant cutoff $k_0$, corresponding to the input PDFs, is used. Note that during the DGLAP evolution, the position of the partons in $b$-space is frozen. Thus such a cascade also forms a hot spot.

Accounting for  the multiple interaction option, that is for contributions containing a few hot spots (that is, a few cascades), we include the eikonal multi-Pomeron contributions, both for the DGLAP and BFKL based descriptions.

Next, we include the enhanced multi-Pomeron diagrams  
introducing the absorption of the low $k_t$ partons. The strength of absorption is driven by the parton density and therefore the effect grows with energy, that is with ln$(1/x)$. We thus have an effective infrared cutoff, $k_{\rm sat}(x)$, which modifies the $k_t$ distribution of the `BFKL' cascade. The result is shown Fig.~\ref{fig:abc}(c), which has some similarity to the DGLAP cascade of Fig.~\ref{fig:abc}(b). However, now the cutoff $k_{\rm sat}$ is not a tuning parameter, but is generated dynamically by the enhanced multi-Pomeron diagrams. Recall that the same diagrams describe high-mass proton dissociation.  That is, the value of the multi-Pomeron vertex simultaneously controls the cross sections of high-mass dissociation and the effective cutoff $k_{\rm sat}$ -- two phenomena which, at first sight, appear to be quite different.

\section{The Durham model}

How may the partonic model of the Pomeron be implemented in practice? To achieve this we note that  the absorption of low $k_t$ partons is 
driven by the opacity, $\Omega$, which depends both on $k_t$ and 
$y=\ln(1/x)$. The opacity, $\Omega_{ik}(y,k_t,b)$, is obtained \cite{KMRnnn} by solving the corresponding BFKL-type
evolution equations in $y$ with a simplified form of the kernel, but which incorporates the main features of BFKL: diffusion in ln$k_t^2$ and $\Delta=\alpha_P^{\rm bare}(0)-1\simeq 0.3$. (A two-channel eikonal is used, $i,k=1,2$.) The inclusion of the $k_t$ dependence is crucial for the transition from the hard to the soft domain. The absorptive factors in the equation embody the result that there is less screening for larger $k_t$. 
The model \cite{KMRnnn} has only a small number of physically motivated parameters, whose values are tuned to reproduce the available high energy $pp$ and $p\bar{p}$ data for $\sigma_{\rm tot}, \;d\sigma_{\rm el}/dt,\; \sigma_{\rm SD}^{{\rm low}M}, \;\sigma_{\rm SD}^{{\rm high}M}/dtdM^2$ etc. Given  $\Omega_{ik}(y,k_t,b)$ we can, in principle, predict all soft and semi-hard inclusive phenomena, such as the survival factors of rapidity gaps, the PDFs and diffractive PDFs at low $x$ and low scales, etc.

It is important to note that hadronization can be incorporated in this partonic description of the Pomeron, via 
Monte Carlo generators, which now would have the advantage of an effective dynamical cutoff $k_{\rm sat}$ to suppress low $k_t$ parton emissions.

In summary, some of the main featuress of the model are

(i) values of the high energy $pp$ total cross section which are suppressed by absorptive corrections. Large values of $\sigma^{{\rm high}M}_{\rm SD}$ enhanced by the increasing phase space with collider energy.

(ii) multi-Pomeron contributions arising from eikonal diagrams, that is the presence of small-size QCD Pomeron cascades (hot spots). This can be tested by measuring Bose-Einstein correlations \cite{BEC}. Specifically, identical pion correlations measure the size of their emission region.

(iii)  multi-Pomeron contributions arising from enhanced diagrams, which lead to the absorption of low $k_t$ partons and automatically introduce an effective cutoff $k_{\rm sat}$ which increases with energy. Due to the cutoff, $k_t>k_{\rm sat}$, the main inelastic process is {\it minijet} production. The dominance of minijets can be tested by observing the two-particle correlations of secondaries at the LHC \cite{KMRmp}.

\section{Implications of latest LHC `soft' data}

Recent measurements at the LHC at 7 TeV are illuminating the soft domain. We give below some general implications, in particular for the Durham model, which follow from the data. The discussion is at a qualitative level, and the observed cross sections are taken at face value, with no attention paid to the (important) experimental errors, simply to illustrate the kind of things that may be learnt from more precise data. 

{\bf Lesson 1}: This concerns the measurements of the inelastic cross section obtained by CMS, ATLAS and ALICE at 7 TeV.  The measured value is defined as the cross section with at least two particles in some central (but far from complete) rapidity, $\eta$, interval.  For instance, ATLAS find 
$\sigma_{\rm inel}=60.3$ mb for the cross section of processes with $M>15.7$ GeV, that is $\xi =M^2/s >5 \times 10^{-6}$ \cite{ATLAS}.  After a model dependent extrapolation to cover the entire rapidity interval they obtain $\sigma_{\rm inel}=69.4$ mb.  CMS find a very similar result, namely 68.0 mb \cite{CMS1}. ALICE also get a similar result \cite{ALICE}.
These estimates are about 5 mb lower than the recent TOTEM value \cite{T}
\begin{equation}
\sigma_{\rm inel}=\sigma_{\rm tot}-\sigma_{\rm el}=73.5 \;{\rm mb}.
\end{equation}
The difference may be attributed to the extrapolated values being 5 mb deficient for low-mass diffraction. (The extrapolation in the high-mass interval is confirmed by the ATLAS measurement $d\sigma/d\Delta\eta \simeq d\sigma/d{\rm ln} M^2 \simeq 1$ mb per unit of rapidity \cite{Newman}.) More specifically, if we define low mass to be $M<2.5$ GeV, then, noting that the unmeasured interval from $M=15.7$ to $M=2.5$ GeV gives $\Delta{\rm ln} M^2 =3.6  $, it follows that the ATLAS, CMS results imply $\sigma_{\rm inel}^{{\rm high}M}\simeq 64$ mb. Then using the TOTEM result we find
\begin{equation}
\sigma_{\rm inel}^{{\rm low}M}\simeq 73.5-64\simeq 10 \;{\rm mb}.
\end{equation}
That is, there should be a considerable amount of low-mass diffraction at 7 TeV; even more than the 7 mb predicted by the Durham model \cite{KMRnnn}.

{\bf Lesson 2}:  As compared to the TOTEM values \cite{T} of $\sigma_{\rm tot}=98.3$ mb and $\sigma_{\rm inel}=73.5$ mb, the Durham model predictions are about 88 mb and 66 mb respectively \cite{KMRnnn}. What does this mean for the Durham model? The model was tuned to describe collider data for $\sigma_{\rm tot}$. At the Tevatron energy, where the CDF and E710 measurements disagree by some 10$\%$, we were much closer to the lower E710 value --- the new data imply we should now tune to a higher value of $\sigma_{\rm tot}$, which will lead to a higher value of the `intercept' $\Delta$ of the hard Pomeron. Secondly, we fitted  to the CERN-ISR estimate of $\sigma_{\rm SD}^{{\rm low}M}=2 $ mb \cite{SD2mb} --- these are the only collider data on low-mass diffraction. The implication is that we should take $\sigma_{\rm SD}^{{\rm low}M}=3 $ mb at $\sqrt{s}=53$ GeV, which is about the maximum permitted by the CERN-ISR data.

{\bf Lesson 3}: We note that the TOTEM value $\sigma_{\rm tot}=98.3$ mb \cite{T} is greater that the original DL value \cite{DL} of $\sigma_{\rm tot}=90.7$ mb at 7 TeV obtained with a soft Pomeron. This indicates that there should be a small-size Pomeron (pQCD Pomeron) contribution which, due to the larger intercept $\alpha_P(0)$. grows faster with energy. Again this implies a higher value of $\Delta$, like 0.35 - 0.4, in the Durham model. 
%The slope $B_{\rm el}=20$ GeV$^{-2}$ is also larger that the 18.5 GeV$^{-2}$ of DL where $\alpha'_P\simeq 0.25\; {\rm GeV}^{-2}$. 

{\bf Lesson 4}: TOTEM find the elastic slope is $B_{\rm el}=20$ GeV$^{-2}$ \cite{T} as compared to the Durham model value of 18.5 obtained with the assumption $\alpha'_P=0$.  The accuracy of the new data indicate that we should re-instate the parameter $\alpha'_P\lapproxeq 0.1\; {\rm GeV}^{-2}$. 

{\bf Lesson 5}: The TOTEM observation that $\sigma_{\rm el}/\sigma_{\rm tot}\simeq 1/4$ implies that the elastic amplitude has essentially saturated at $b=0$.

{\bf Lesson 6}: Of course, the Durham model was not meant to be applicable for $|t|$ values beyond the forward peak in elastic scattering. It was designed to describe the main features of soft and semi-hard inclusive processes in terms of a partonic approach, and not rare exclusive large $|t|$ processes. But we may ask if it would be possible to reproduce the dip seen by TOTEM in $pp$ elastic scattering at $-t=0.53$ GeV$^{2}$ \cite{T0}. This could be done, but would require a many-channel eikonal model (that is more Good-Walker diffractive eigenstates) with correspondingly more parameters.

{\bf Lesson 7}: The difference in $t$ shape between the $pp$ cross section at 7 TeV (dip) \cite{T0} and the $p{\bar p}$ at the Tevatron energy (shoulder starting at $-t\simeq 0.6$ GeV${^2}$ \cite{D0}) may be thought to indicate the presence of odderon exchange. However the comparison of the CERN-ISR $pp$ and $p{\bar p}$ elastic data \cite{ISR} give a much stronger limit on the odderon amplitude. These data imply no visible odderon effects in $d\sigma_{\rm el}/dt$ at the LHC (assuming almost no energy dependence, $\alpha_{\rm odd}(0)\simeq 1$, and a rather flat $t$ behaviour of the odderon).

{\bf Lesson 8}: The approximate TOTEM behaviour $d\sigma_{\rm el}/dt \propto 1/t^8$  at larger $|t|$ \cite{T0} cannot be due to {\it asymptotic} three gluon exchange between three pairs of quarks since we see no $1/t^8$ behaviour at the CERN-ISR in the same $t$ interval.

\section*{Acknowledgements} We thank Carlos Merino and Christophe Royon for arranging such an enjoyable Workshop.

\end{document}